\documentclass[aps,pra,reprint,superscriptaddress,amsmath,amssymb,floatfix]{revtex4-2}
\usepackage[english]{babel}

\babelprovide[import, main]{english}

\usepackage{graphicx}          
\usepackage{dcolumn}           
\usepackage{bm}                
\usepackage{hyperref}          
\usepackage{xcolor}            
\usepackage{amsmath}           
\usepackage{physics}           
\usepackage{siunitx}           
\usepackage{cleveref}
\hypersetup{
  colorlinks = true,
  linkcolor  = blue!70!black,
  citecolor  = blue!70!black,
  urlcolor   = blue!70!black,
}

\newcommand{\BFSR}{Bethe $f$-sum rule}

\newcommand{\beq}{\begin{equation}}
\newcommand{\eeq}{\end{equation}}
\newcommand{\bea}{\begin{eqnarray}}
\newcommand{\eea}{\end{eqnarray}}

\newcommand{\ofqw}{(q,\omega)}

\newcommand{\qv}{{\bf q}}

\newcommand{\e}{\mathrm{e}}

\newcommand{\mbf}[1]{\mathbf{#1}}

\newcommand{\invangs}{\mathrm{\AA}^{-1}}
\renewcommand{\vec}[1]{\mathbf{#1}}

\usepackage{booktabs}
\usepackage{tikz}
\usepackage{pgfplots}

\pgfplotsset{compat=1.18} 

\begin{document}

\title{Reconciling chemical models of X-ray Thomson Scattering with the Bethe $f$-sum rule}

\author{Maximilian P.~B\"ohme}
\email{boehme4@llnl.gov}
\affiliation{%
 Lawrence Livermore National Laboratory, Livermore, California, 94550, USA
}

\author{Paul Hamann}
\affiliation{%
 Lawrence Livermore National Laboratory, Livermore, California, 94550, USA
}
\affiliation{%
Institute of Radiation Physics, Helmholtz-Zentrum Dresden-Rossendorf (HZDR), D-01328 Dresden, Germany
}
\affiliation{%
Institut f\"ur Physik, Universit\"at Rostock, D-18057 Rostock, Germany
}

\author{Veronika A.~Kruse}
\affiliation{%
 Lawrence Livermore National Laboratory, Livermore, California, 94550, USA
}
\author{Hannah~M.~Bellenbaum}
\affiliation{%
Center for Advanced Systems Understanding (CASUS), D-02826 G\"orlitz, Germany
}
\affiliation{%
Institut f\"ur Physik, Universit\"at Rostock, D-18057 Rostock, Germany
}

\author{Armin Bergermann}
\affiliation{%
SLAC National Accelerator Laboratory, Menlo Park CA 94309, USA
}

\author{David T.~Bishel}
\affiliation{%
 Lawrence Livermore National Laboratory, Livermore, California, 94550, USA
}

\author{Thomas Gawne}
\affiliation{%
Center for Advanced Systems Understanding (CASUS), D-02826 G\"orlitz, Germany
}

\author{Dirk~O.~Gericke}
\affiliation{Centre for Fusion, Space and Astrophysics, Department of Physics, University of Warwick, Coventry CV4 7AL, United Kingdom}

\author{Zhandos~A.~Moldabekov}
\affiliation{%
Institute of Radiation Physics, Helmholtz-Zentrum Dresden-Rossendorf (HZDR), D-01328 Dresden, Germany
}

\author{Pontus Svensson}
\affiliation{%
Institute of Radiation Physics, Helmholtz-Zentrum Dresden-Rossendorf (HZDR), D-01328 Dresden, Germany
}

\author{Jan Vorberger}
\affiliation{%
Institute of Radiation Physics, Helmholtz-Zentrum Dresden-Rossendorf (HZDR), D-01328 Dresden, Germany
}
\author{Tobias Dornheim}
\affiliation{%
Institute of Radiation Physics, Helmholtz-Zentrum Dresden-Rossendorf (HZDR), D-01328 Dresden, Germany
}
\affiliation{%
Center for Advanced Systems Understanding (CASUS), D-02826 G\"orlitz, Germany
}
 
\date{\today}

\begin{abstract}
X-ray Thomson scattering (XRTS) is a key diagnostic for high-energy-density plasmas, which can exhibit significant quantum effects even at elevated temperatures. XRTS experiments are commonly interpreted using the Chihara decomposition, that was derived in the chemical picture and, thus, separates contributions from bound and free electrons. Despite being the de-facto standard for analysing measurements, a well-known shortcoming is that the standard bound-state treatment in the form of the impulse approximation fails to satisfy fundamental theoretical constraints, most notably the Bethe $f$-sum rule (BFSR). The problem arises due to the usage of plane waves in the impulse approximation as well as non-negligible contributions from bound-bound transitions. In this work, we present an analytical extension of the Chihara decomposition of the dynamic structure factor for matter in the ground state, using hydrogenic bound-free and bound-bound transitions. We demonstrate that compliance with the BFSR is only achieved when bound-bound transitions are explicitly included {\em and} an exact treatment of the bound-free contribution is applied. Therefore, the analytic model solves an important issue that has plagued the interpretation of XRTS experiments and represents the ground work for future finite temperature extensions. Finally, detector ray-tracing simulations for atomic hydrogen demonstrate experimentally detectable deviations from the standard Chihara model. The model will be made available in the open source XRTS library \texttt{xDAVE} [\textit{Bellenbaum et al.}, Phys. Plasmas \textbf{33}, 083904 (2026)].
\end{abstract}


\maketitle

\section{\label{sec:intro}Introduction}
X-ray Thomson Scattering (XRTS) experiments have proven to be one of the most essential tools in order to understand warm dense matter (WDM) \cite{glenzer_x-ray_2009,vorberger_roadmap_2025,dornheim2026overviewxraythomsonscattering}, as they provide the ability to gain significant insight into a plethora of effects such as miscibility~\cite{wunsch_x-ray_2011,Frydrych2020}, plasmon damping~\cite{Neumayer_PRL_2010,Sperling_PRL_2015,bespalov_momentum-resolved_2026}, and effective ionization~\cite{Gregori_PRE_2003,doppner_observing_2023}.
This knowledge is especially useful as WDM sits in a ``anti-Goldilocks" zone in the plasma phase diagram. It is characterized by several eVs of temperature, pressures up to $10^4$ GPa and number densities ranging form $10^{21} - 10^{25}$ cm$^{-3}$. Thus, it can neither be accurately captured by models that assume an ideal plasma phase nor a dense plasma phase with perfect screening. In general, one must rely on \textit{ab initio} simulation techniques such as density functional theory (DFT) \cite{Bergermann2026a, Desjarlais2004, moldabekov_enhancing_2026,Baczewski_PRL_2016} or path-integral Monte Carlo (PIMC) \cite{dornheim_unraveling_2025,Boehme_PRL_2022}. While these simulation techniques offer a high accuracy, they have the drawbacks of being computationally expensive and sometimes being not straightforward to interpret~\cite{bellenbaum_estimating_2025}. 
Although plentiful progress has been reported over the last decade~\cite{Baczewski_PRL_2016,dornheim_unraveling_2025,Dornheim_POP_2025,moldabekov_enhancing_2026,schorner_x-ray_2023,bespalov_momentum-resolved_2026}, \textit{ab initio} simulations still do not constitute the most widely used method for the interpretation of experimental measurements~\cite{dornheim2026overviewxraythomsonscattering}.
It is therefore desirable to first treat these spectra using more approximate analytical models to obtain a rough estimate of the measured conditions. Once the possible parameter space has been reduced, one can utilize \textit{ab initio} techniques such as linear-response time-dependent DFT (LR-TDDFT) \cite{bespalov_momentum-resolved_2026} to obtain a highly accurate interpretation of the experiment. Furthermore, simple models based on a single-particle theory provide a degree of interpretability as one can distinguish the contributions of strongly bound core electrons, weakly bound valence states and free electrons -- a distinction that is generally lost in a proper many-body description.


XRTS probes equilibrium electron density fluctuations in the target and is therefore governed by the electronic dynamic structure factor (DSF) \cite{chapman_analysis_2012,chapman_observation_2015,doppner_observing_2023}. One highly popular model of computing the DSF of WDM is the so-called Chihara decomposition \cite{chihara_difference_1987,chihara_interaction_2000,boehme2023evidence}. This was the first method used together with XRTS for diagnosing warm dense matter and is still the most used theory to interpret XRTS experiments at X-ray free electron laser facilities (XFEL) \cite{fletcher_ultrabright_2015} as well as inertial confinement fusion facilities \cite{doppner_observing_2023,kritcher_measurement_2020}; see Ref.~\cite{dornheim2026overviewxraythomsonscattering} for a recent overview of more than 90 WDM XRTS experiments. Several works have used the Chihara decomposition for synthetic data generation \cite{poole_case_2022,poole_multimessenger_2024}, comparison against PIMC simulations \cite{bellenbaum_estimating_2025} and in conjunction with feasibility studies regarding the physics of spectrometers~\cite{gawne_heart_2026,gawne_effects_2024}.


While both \textit{ab initio} simulations and forward fitting to Chihara models require some assumptions, recent advancements of XRTS in combination with imaginary-time formalism methods have established exact relations on how to diagnose temperature \cite{dornheim_accurate_2022,gawne2026modelfreeinterpretationxraythomson} under the knowledge of the source and instrument function (SIF) \cite{gawne_effects_2024}. Not only does this technique offer a clear way of uncertainty quantification \cite{dornheim_imaginary-time_2023}, it can also be used to calculate the electronic static structure factor~\cite{dornheim_x-ray_2024}, Rayleigh weight~\cite{Dornheim_POP_2025}, and static linear density response~\cite{schwalbe2025staticlineardensityresponse} without any physics approximations. All of this can be used as a complete workflow to fully analyse the plasma state \cite{bohme_correlation_2026} of an HED experiment. Chihara decomposition models would be an excellent tool to drive the development of imaginary-time based inference techniques as they offer a fast and simple way to calculate a dynamic structure factor suitable to approximately describing the plasma environment. However, the application of imaginary-time correlation function based techniques requires the adherence of the DSF to exact relations, a condition that is currently not met due to the violation of the {\BFSR}.

Over the years, the Chihara decomposition has been refined in several ways, including the use of hyper-netted-chain (HNC) calculations for the elastic scattering component to account for ionic structure in the plasma \cite{wunsch_ion_2009,wunsch_x-ray_2011}, as well as average-atom treatments of bound-free transitions \cite{johnson_thomson_2012}. A major criticism of practical implementations of the Chihara model is that the usual treatment of transitions from and into bound states, via the impulse approximation (IA) for example, can violate the {\BFSR}. This violation has been documented in Ref.~\cite{gregori_electronic_2004} and introduces uncertainties in the normalization of the bound-free contribution. It affects not only the accuracy of XRTS modelling codes, but also the use of techniques that rely on the Bethe $f$-sum rule to be satisfied, such as the normalization of imaginary-time correlation functions \cite{dornheim_x-ray_2024,bellenbaum_toward_2025,bellenbaum_estimating_2025}. More broadly, this sum rule has also been used in other contexts, for example to establish an absolute scale for DSF measurements in synchrotron experiments \cite{schulke_dynamic_1995} or to assess the quality of theoretical models~\cite{chuna2026merminsdielectricfunctionfsum}. Here, we demonstrate that the origin of this violation in current codes \cite{chapman_simulating_2014,bellenbaum_x-ray_2026,lutgert_jaxrts_2026} lies in two approximations: the neglect of bound-bound transitions, as previously predicted in Ref.~\cite{baczewski_predictions_2021}, and the use of the impulse approximation for bound-free transitions~\cite{eisenberger_compton_1970}. In his original formulation, Chihara~\cite{chihara_interaction_2000} introduced a formally correct bound-electron DSF that can, in principle, include bound-bound transitions. However, in practical implementations of the Chihara decomposition used to analyse XRTS spectra, this contribution is commonly neglected. We show that including bound-bound transitions and treating bound-free transitions exactly can have a significant impact on simulated XRTS spectra. Conversely, describing these contributions within a highly approximate framework can, for example, lead to the misidentification of bound-free features.

The paper is organized as follows. In \Cref{sec:theory}, we review the Bethe $f$-sum rule, and the TRK sum rule as its long-wavelength limit, along with a section dedicated to the Chihara decomposition. This analysis shows that bound-bound transitions must be included in order to satisfy the Bethe $f$-sum rule. We then present a minimal bound-state extension of the Chihara model for ground-state hydrogen. The new bound-bound and bound-free models are described in the main text, while details of the calculations are given in Appendices~\cref{app:bf_numerical,app:bound-free,app:matrix}. The resulting bound-free DSFs are compared to those obtained from the commonly used impulse approximation. In \Cref{sec:results}, we show that the Bethe $f$-sum rule is satisfied exactly only when both the exact bound-bound and bound-free expressions are used. As a final comparison, we simulate detector images for ground-state atomic hydrogen using ray-tracing simulations. In \Cref{sec:conclusion}, we give an outlook on how the model can be extended to finite-temperature systems to improve the accuracy of Chihara models in the interpretation of XRTS experiments.

\section{\label{sec:theory}Theory}

A typical XRTS setup consists of an X-ray source that illuminates a target with an incident beam of wavevector $\mathbf{q}_i$ and frequency $\omega_i$.
Scattered photons are collected by a detector at a fixed angle, from which the scattered wavevector $\mathbf{q}_s$ and frequency $\omega_s$ are determined.
The momentum and energy transfers to the system are then $\mathbf{q} = \mathbf{q}_s - \mathbf{q}_i$ and $\omega = \omega_s - \omega_i$, respectively.
The double differential cross-section for this process is given by~\cite{chihara_difference_1987,chihara_interaction_2000,glenzer_x-ray_2009}
\begin{equation}
    \frac{\partial^2 \sigma}{\partial \Omega \, \partial \omega}(\mathbf{q},\omega)
    = \sigma_T \frac{q_s}{q_i} S_{ee}(\mathbf{q}, \omega),
\end{equation}
where $\sigma_T = 6.65 \times 10^{-29}$~m$^2$ is the Thomson cross-section, the factor $q_s/q_i$ accounts for the photon phase-space kinematics, and $S_{ee}(\mathbf{q},\omega)$ is the electronic DSF of the system. Thus, the physics probed by the X-ray beam is completely contained inside the electronic DSF of the system.

\subsection{Bethe $f$-sum rule} 

We start with an overview of the {\BFSR}, as it is one of the basic sum-rules that must be fulfilled in general for any electronic density-density DSF.

The many-body DSF in equilibrium for a system consisting of $N$ identical particles is 
given by \cite{giuliani_quantum_2008,schulke2007electron} 
\begin{equation}
    S(\mbf{q},\omega) = \sum_{I} P_I(T) \sum_{F} |\bra{F} \sum_{j=1}^{N} e^{i\mbf{q}\hat{\mbf{r}}_j} \ket{I}|^2 \delta(\omega - \varepsilon_F + \varepsilon_I),
\end{equation}
with $I,F$ denoting all initial and final many-body states, $P_I(T)$ the initial occupation probability as a function of the system temperature $T$, and $\varepsilon_{I/F}$ denoting the initial and final eigenenergy of the corresponding many-body state, respectively. 
Based on the work of Thomas, Reiche and Kuhn \cite{reiche_uber_1925,kuhn_uber_1925}, Bethe showed~\cite{bethe_bremsformel_1932} that the first frequency moment of the DSF is given by
\begin{equation} \label{eq:fsum_general}
    \int \mathrm{d}\omega \, \omega \, S(\mbf{q},\omega) = N \frac{\mbf{q}^2}{2},
\end{equation}
which is generally referred to as Bethe $f$-sum rule. It can be derived through the continuity equation in the general many-body case and remarks that electrons respond to perturbations at high frequencies as individual particles~\cite{giuliani_quantum_2008}. 

The DSF of a single electron in an initial state $\ket{i}$ is given by \cite{mattern_theoretical_2013,schulke2007electron}
\begin{equation} \label{eq:DSF_GS_i} 
    S_\text{i}(\qv,\omega) = \sum_{f} |\bra{f}e^{i\qv \mbf{r}} \ket{i}|^2 \, \delta(\omega - \varepsilon_f + \varepsilon_i).
\end{equation}
Here, $\varepsilon_{f,i}$ are the energies of the final and initial state. 
Summing over all possible final states $\bra{f}$ states gives a contribution for each transition $i\to f$, with a weight given by the plane-wave matrix element
\begin{equation}
M_{fi}(\qv)= \bra{f} e^{i \mbf{q} \mbf{r}} \ket{i}.
\end{equation}

In this case, Eq.~(\ref{eq:fsum_general}) for the first frequency moment $\Omega^{(1)}$ of the DSF becomes 
\begin{equation}\label{eq:f_sum_rule}
    \Omega^{(1)}(\mbf{q}) = \int \mathrm{d}\omega \ \omega \, S_i(\qv,\omega) = \frac{\qv^2}{2}\,.
\end{equation}
This requirement is met exactly when the sum over final states in 
Eq.~\eqref{eq:DSF_GS_i} is taken over a complete basis set $\{\ket{f}\}$ of a Hilbert space such that the Hamiltonian of the system $\hat{H}$ can be represented by 
 \begin{equation}
     \sum_f \varepsilon_f \ket{f} \bra{f} = \hat{H}.
 \end{equation}
For a single electron in an  attractive Coulomb potential, which is the model under consideration, the final state sum must therefore include both bound states, that is, states with negative energy eigenvalues 
$\varepsilon_{n\ell m}$,
indexed by discrete quantum numbers $n,\ell,m$, and the continuum of positive energy (scattering) states.

\subsection{The Chihara decomposition}

The electronic DSF in the Chihara decomposition \cite{chihara_difference_1987,chihara_interaction_2000,siegfried_review} is commonly denoted as
\begin{equation}
    Z_\text{tot}S_{ee}^\text{tot}(\qv,\omega) = W_R(\qv) \delta(\omega) + S_{ee}^\text{inel.}(\qv,\omega),
\end{equation}
by separating the spectrum into an elastic contribution, the Rayleigh weight $W_R$, and the inelastic part $S_{ee}^{inel.}$. The Rayleigh weight accounts for contributions from electrons associated with nuclei (electrons being either in bound states or in the screening cloud) due to elastic scattering~\cite{dornheim_model-free_2025}. $Z_\text{tot}$ denotes the total nuclear charge. Formally, the DSF is a function of the momentum transfer vector $\qv$ and the frequency $\omega$. Since we are interested in applications to homogenous and isotropic systems, one can show that the DSF will only depend on the magnitude of the momentum transfer vector $q = |\qv|$. For a more thorough review of the scattering kinematic of X-rays on dense plasmas we refer the readers to Refs.~\cite{bellenbaum_x-ray_2026,glenzer_x-ray_2009,bohme_correlation_2026}.

Here, we focus on the inelastic part of the spectrum that is usually split into the inelastic scattering on free electrons and on bound electrons
\begin{equation}\label{eq:sffbfbb}
    S^\text{inel}_{ee}\ofqw = S^\text{ff}_{ee}\ofqw + S^\text{bf}_{ee}\ofqw + S^\text{bb}_{ee}\ofqw.
\end{equation}
The treatment of inelastic scattering of bound electrons only through bound-free scattering [2nd term in Eq.(\ref{eq:sffbfbb})] is an approximation employed in all common XRTS codes, while the full treatment of inelastic scattering of bound electrons also includes transitions from a bound state to another bound state [3rd term in Eq.(\ref{eq:sffbfbb})]. We show that this approximation is primarily at fault for the violation of the {\BFSR}.

For a plasma of the average charge-state $Z$, the inelastic free-free scattering is given by 
\begin{equation}
    S^\text{ff}_{ee}\ofqw = Z_f S_{ee}^{0}\ofqw, 
\end{equation}
where $Z_f$ is the number of free electrons and $S^{0}_{ee}\ofqw$ is usually taken to be the DSF of the uniform electron gas \cite{dornheim_effective_2020,dornheim_dynamic}, potentially also taking electron-ion collisions into account \cite{glenzer_observations_2007,Neumayer2010}. Since our primary focus here is on the inelastic treatment of the bound electrons, we will restrict ourselves to the single-atom ground state, where per definition the system  will not have any excitations and thus the scattering on free electrons must vanish.

\subsection{Violation of the Bethe $f$-sum rule in the impulse approximation}
\begin{figure}
    \centering
    \includegraphics[width=\columnwidth]{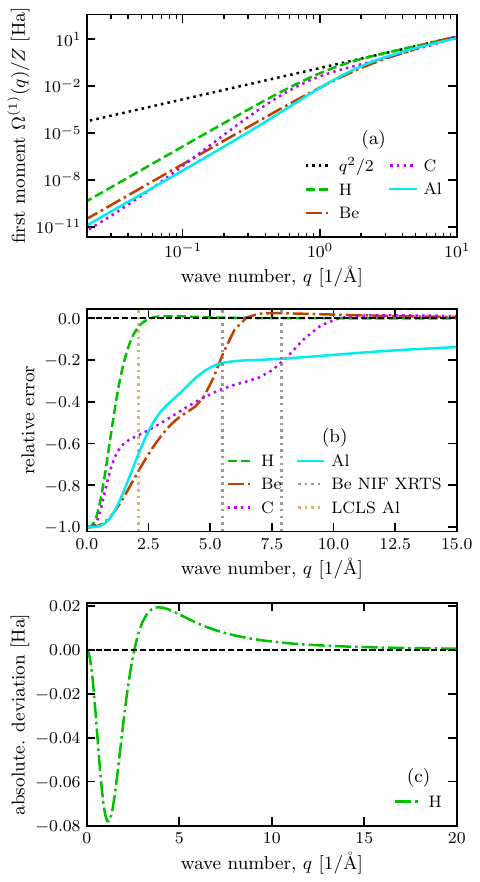}
    \caption{Violation of the {\BFSR} in the impulse approximation (IA). Panel (a) depicts the first frequency moment of the IA divided by the nuclear charge state for the elements H (green), Be (red), C (magenta) and Al (turquoise) as a function of the wave number compared to the analytical dispersion relation of the {\BFSR} (black dashed). Panel (b) illustrates the relative error of the $f$-sum rule as a function of the wave-number, where the gray vertical lines depicts the accessible wave-number in Refs.~\cite{doppner_observing_2023,dornheim_unraveling_2025}. The gold dashed vertical line depicts maximum wave number for the XRTS experiments on Aluminum in Ref.~\cite{Sperling_PRL_2015}. Panel (c) shows the absolute deviation from the $f$-sum rule for H. }
    \label{fig:violation_higher_Z}
\end{figure}

\Cref{fig:violation_higher_Z} illustrates the physical setting and the central question addressed in this work. Panel (a) shows the first frequency moment $\Omega^{(1)}(q)$, defined in \Cref{eq:f_sum_rule}, per nuclear charge $Z$ for the bound-free contribution in the IA \cite{schumacher_incoherent_1975,eisenberger_compton_1970}. The results are calculated for ground-state H, Be, C, and Al using the implementation in the \texttt{xDAVE} code \cite{bellenbaum_x-ray_2026}. The black dashed line indicates the quadratic dispersion required by the Bethe $f$-sum rule. The IA is seen to violate the $f$-sum rule most strongly at small $q \lessapprox 1.0 \, \mathrm{\AA}^{-1}$, while approaching the correct asymptotic limit at large $q$. 

Panel (b) of \Cref{fig:violation_higher_Z} further quantifies the violation by showing the corresponding relative deviations. The discrepancy between the IA and the Bethe $f$-sum rule depends strongly on the wave number and increases in range with nuclear charge. For hydrogen, the IA approaches the correct dispersion at approximately $2.5\,\mathrm{\AA}^{-1}$, whereas beryllium requires wave numbers of about $5\,\mathrm{\AA}^{-1}$. For carbon and aluminum, the deviations extend into wave-number ranges relevant to typical backscattering XRTS experiments; carbon reaches the correct limit only near $10\,\mathrm{\AA}^{-1}$. The aluminum deviation persists over a substantially wider range spanning the whole experimentally accessible regime. 
Experimentally accessible wave numbers are shown in form of the vertical gray dashed lines for the Beryllium shots at NIF from Refs.~\cite{doppner_observing_2023,dornheim_unraveling_2025} at $7.9$ $\mathrm{\AA
}^{-1}$ and $5.5$ $\mathrm{\AA}^{-1}$, respectively. The dark-yellow vertical dashed line depicts the largest wave number reported in Ref.~\cite{Sperling_PRL_2015} for shots taken at the Linac Coherent Light Source (LCLS) on Aluminum at $2.1$ $\invangs$. The present ground-state example therefore suggests that current XRTS interpretation methods may be insufficiently accurate, as they rely exclusively on the IA to describe bound-state transitions, leading to significant violations of the {\BFSR} at the experimentally accessible wave numbers.

Panel (c) of \Cref{fig:violation_higher_Z} shows as an example the absolute deviation of the IA first frequency moment for hydrogen from the $f$-sum rule. An interesting development here is that the IA moment first underestimates the $f$-sum rule up until $\approx 2.5$ $\mathrm{\AA}^{-1}$ and then starts to overestimate the analytical result while slowly converging towards it. Even while the relative violation may be close to zero for $q \approx 2.5 $ $\mathrm{\AA}^{-1}$, the IA still violates the $f$-sum rule for larger $q$.  

While one can argue that this problem primarily affects the ground state, it has not yet been proven that this violation of the {\BFSR} is not problematic in an HED setting. Since the IA or some of its corrected forms are the primary method to treat bound electron DSFs, one may miss crucial electronic transitions relevant to interpret experimental data. 

The violation has been discussed in previous work by Mattern and Seidler~\cite{mattern_theoretical_2013}, who give a reason why the IA does not preserve the {\BFSR}, due to its inconsistent treatment of eigenstates, but do not include any bound-bound transitions. Other works have already used the Chihara decomposition to calculate DSFs for XRTS in HED setting from average atom type orbitals \cite{johnson_thomson_2012} that may also not fulfil the {\BFSR} due to the neglect of bound-bound transitions, while treating the bound-free transitions coherently. Codes such as FEFF9 \cite{rehr_parameter-free_2010} have been used to generate the real-space Green's functions approach in Ref.~\cite{mattern_theoretical_2013}, but have unfortunately never been widely adapted for the analysis of XRTS experiments. The presence of bound-bound transitions in WDM has previously been reported in Ref.~\cite{baczewski_predictions_2021}.

Historically, simplified Chihara models were a well-justified starting point \cite{glenzer_x-ray_2009}: both drive lasers, probe beams and detectors lacked the precision and photon energy range needed to resolve the fine spectral features that a more rigorous treatment would affect. Notably, the theoretical requirements for satisfying the {\BFSR} have long been established in the synchrotron scattering community, but this knowledge did not carry over into the XRTS field. The advent of facilities such as the European XFEL fundamentally changes this situation, demanding a more rigorous treatment if XRTS experiments are to be interpreted with high fidelity. 


\subsection{Hydrogen-like bound-free dynamic structure factor}
We can write down the bound-free contribution to the DSF incorporating transitions from the bound state $(n,\ell,m)$ to the continuum of free states as follows
\begin{equation}\label{eq:dsf_bf_nlm}
    S_{n\ell m}^{bf}(\qv,\omega) = \int \mathrm{d}^3k \, |M^{bf}_{\mathbf{k},n\ell m}(\mbf{q})|^2 \delta(\omega - \varepsilon_\mbf{k} + \varepsilon_{n\ell m}),  
\end{equation}
with $\mbf{k}$ as the final state wave-vector for the continuum  with the corresponding single-particle continuum energy contribution $\varepsilon_\mbf{k}= k^2/2$.
In this case, the transition matrix element takes the following form
\begin{equation}\label{eq:M_k_nlm}
    M^{bf}_{\mbf{k},n\ell m}(\mbf{q}) = \int\,\mathrm{d}^3r \, {\psi_{\mbf{k}}^\text{free}}^*(\mbf{r}) e^{i\mbf{qr}} \psi^\text{bound}_{n\ell m}(\mbf{r}).
\end{equation}
For the analysis of XRTS experiments, this matrix element has been commonly evaluated in the IA \cite{eisenberger_compton_1970,schumacher_incoherent_1975} which is obtained
after approximating the continuum wavefunction by its asymptote given by a plane wave, i.e. setting $\psi^\text{free}_\mbf{k}(\mbf{r}) \approx e^{i\mbf{k}\mbf{r}}$ and evaluating the integral in the limit of large momentum, see App.~\ref{app:impulse}.

However, in order to fulfil the $f$-sum rule to arbitrary precision, the bound-free matrix element needs to be treated consistently with the full eigenbasis of the hydrogen Hamiltonian. To this end, Eq.~\eqref{eq:M_k_nlm} needs to be evaluated using the actual Coulomb continuum wavefunction, given by \cite{AbramowitzStegun1964,pollock_properties_1988}
\begin{equation}\label{eq:continuum}
    \Psi_{\mbf{k}}(\mbf{r}) = \Gamma(1-i\eta)e^{-\pi\eta/2} e^{i\mbf{kr}} {}_1F_1(i\eta,1,-ikr - i \mbf{k} \cdot \mbf{r}),
\end{equation}
with ${}_1F_1(a,b,z)$ denoting the confluent hypergeometric function. 

We give a derivation of both a numerical and analytical solution in App.~\ref{app:matrix} and \ref{app:bound-free}, respectively. The analytical solution for the matrix element is well known~\cite{belkic_bound-free_1981,omidvar_ionization_1972,moses_bounds_1994}. The solution is based on the Nordsieck's analytical solution of the Bremsstrahlungs-integral \cite{nordsieck_reduction_1954} and Bethe and Maximon's idea to use it as a generating function for the matrix elements \cite{bethe_theory_1954}.

An important realization is that the analytical solution of the bound-free matrix element in \Cref{eq:bound_free_analytic} is only dependent on the polar direction $\cos(\vartheta) = \mu$. 
When carrying out the integral Eq.~\eqref{eq:dsf_bf_nlm} in spherical coordinates, due to the delta distribution, the integrand vanishes unless $|\mbf{k}| = \sqrt{2(\omega - \varepsilon_{n\ell m})}$. The remaining azimuthal integral can be carried out numerically using Gauss-Kronrod quadrature nodes. The final expression reads
\begin{equation}
    S^{bf}_{n\ell m}\ofqw = 2\pi \nu \int_{-1}^{1}\, \mathrm{d}\mu \, \mu\, |M^\text{bf} _{\mbf{k}, n\ell m}(q) |^2,
\end{equation}
where $\nu= \sqrt{2(\omega - \varepsilon_{nlm})}$ and $\mbf{k}=\nu\mu \hat{e}_z$.

We can therefore get the exact ground-state DSF of hydrogen-like atoms. This ansatz will serve as a basis for future finite-temperature XRTS bound-free models, that gives one a high flexibility and also apply it to current experimental platforms.

\begin{figure}[tbp]
  \centering
  \includegraphics[width=\columnwidth]{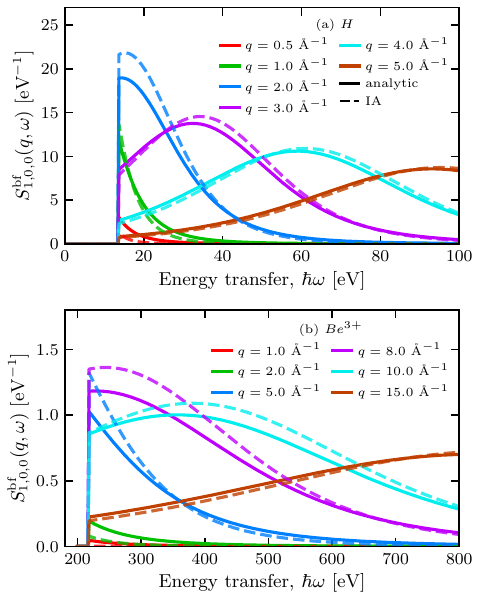}
  \caption{\label{fig:ia_vs_full_bf}%
    DSF using the Schumacher Impulse approximation (dashed) and the full Coulomb wave function (solid) for several values of $q$ as a function of the frequency in $\textrm{eV}^{-1}$. Panel (a) depicts the comparison for $H$, panel (b) for $Be^{3+}$.
  }
\end{figure}

A comparison between the bound-free DSF in the IA (dashed) and the analytical hydrogen model (solid) is shown in \Cref{fig:ia_vs_full_bf} panel (a) for hydrogen at $q$'s between $0.5 $ $\mathrm{\AA}^{-1}$ and $5.0$ $\mathrm{\AA}^{-1}$. The result is as expected that the IA is in reasonable agreement at $q = 4.0$ $\invangs$. However, for small $q < 4.0$ $\invangs$  the deviations become more dominant. As discussed, this is due to the incorrect treatment of the continuum Coulomb states by approximating them as a plane waves. Overall, we see that the IA becomes a reasonable approximation with increasing $q$. However, the point where the IA becomes precise enough is highly dependent on the element in question. Panel (b) depicts the comparison for H-like Be for various $q$. Here only the highest $q = 15$ $\invangs$ reasonably agrees with the analytic solution. Therefore it reveals a practical issue regarding the accuracy of the IA for the interpretation of XRTS experiments on even low-Z elements. The IA deviates from the exact analytic result at $T=0\,\mathrm{K}$ for experimentally accessible $q$. Extending the theory to finite-$T$ cases therefore remains an important direction for future work, since most WDM experiments operate in this regime. The current work, however, sheds light on the accuracy of using only the IA to treat inelastic bound transitions in WDM.

\subsection{Hydrogen-like bound-bound dynamic structure factor}

The bound-bound DSF for an electron in the hydrogen-like state $\ket{i} = \ket{n,\ell,m}$ is given by a sum over all possible final bound states $\ket{f} = \ket{n',\ell',m'}$ 
\begin{equation}\label{eq:DSF_bb_general_case}
    S_{i}^{bb}\ofqw = \sum_{f} |M^{bb}_{fi}(\mbf{q})|^2 \delta(\omega - \varepsilon_f + \varepsilon_i),
\end{equation}
The evaluation of the matrix element is therefore the only practical difficulty and is given by
\begin{equation} \label{eq:bound-bound_matrix_element_general}
    M^{bb}_{fi}(\mbf{q}) = \int d^3r\, {\psi^{\text{bound}}_{n'\ell'm'}}^\ast (\mbf{r}) e^{i\mbf{qr}} \psi^\text{bound}_{n\ell m}(\mbf{r}).
\end{equation}
While it is in principal possible to compute this the matrix element numerically in spherical coordinates (see App.~\ref{app:matrix}), doing so may become unfeasible if a large number of transitions need to be taken into account. It is much more practical to evaluate \Cref{eq:bound-bound_matrix_element_general} analytically by the introduction of parabolic coordinates as described in Appendix~\ref{app:bound-bound}.
    
Fig.~\ref{fig:M_bb} shows the absolute value of the bound-bound matrix element depending on the wave number $q$ for selected transitions. In agreement with Eq.~\eqref{eq:dipole_limit}, at $q=0$ all amplitudes vanish, unless the initial and final state are the same, in which case $|M_{fi}| = 1$. Transitions which are \textit{dipole-allowed} approach linear behaviour towards small $q$, with a rate given by the dipole matrix element as shown by the dotted black lines. For the initial state being the $1s$ orbital (upper panel) transitions to $p$-orbitals are dipole-allowed. In all other cases, the first derivative of $M(q)$ vanishes at $q=0$. Matching the nodal structure of the $1s$-orbital, for each transition there is a single $q$ of maximum amplitude, from which on $|M_{fi}|$ monotonously decays and vanishes as $q \to \infty$. 

The situation is different in the bottom panel (b) which shows results for an initial $2p$-state. Here, transitions to $s$- or $d$-orbitals are \textit{dipole-allowed}. Depending on the final state, $|M_{fi}|$ possesses a different number of local maxima with zeros in between. When directly considering $M_{fi}$ instead of its absolute value, which depending on the transition is a purely real or purely imaginary quantity, we can see that the real (or respectively imaginary) part oscillates around zero. 

\begin{figure}[htbp]
  \centering
  \includegraphics[width=\columnwidth]{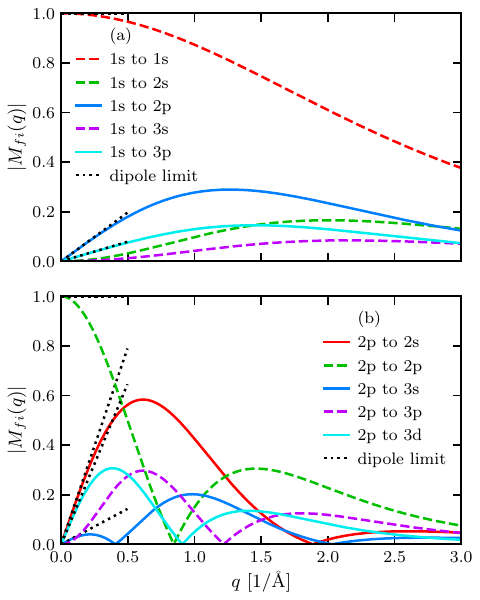}
      \caption{\label{fig:M_bb} Absolute value of scattering matrix elements $M_{fi}$ for different bound-bound transitions. Panel (a) depicts the transitions for the 1s state into various states. The black dashed lines depict the dipole limit for allowed bound-bound transitions. Panel (b) shows the transition matrix elements originating from the 2p state.}
\end{figure}
The delta function in \Cref{eq:DSF_bb_general_case} enforces energy conservation of the scattering process. However, the initial and final states elastically scatter over a finite bandwidth due to the natural lifetime of the states \cite{Cowan1981}, Doppler shift due to thermal motion of the ions \cite{Griem2005}, and the perturbation of the state eigenenergies by the plasma environment \cite{Volonte1978, More1982, Li2012}. A thorough investigation of the line-shape function for bound-bound transitions, particularly in dense plasmas, is a detailed pursuit in its own right \cite{Gomez2022}.

\section{Results} \label{sec:results}
\subsection{Bethe $f$-sum rule}
\begin{figure}
  \centering
  \includegraphics[width=\columnwidth]{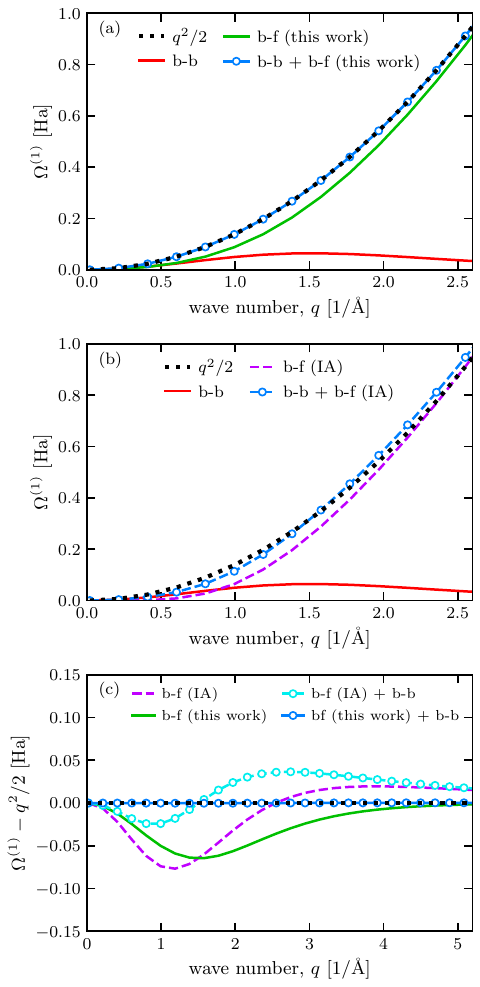}
  \caption{\label{fig:bethe_fulfilled}%
    Contributions to the first moment of $S(q,\omega)$ for an electron in the $1s$-orbital. Panel (a) shows the analytic result (black dashed), the bound-bound contribution (red) and the bound-free contribution (green). Panel (b) replace the exact bound-free treatment with the IA. Panel (c) shows the absolute deviation of the summed up components (circles) for the two cases and the bound-free treatment only.}
\end{figure}

We now complete the discussion by showing that the analytic models indeed fulfil the Bethe $f$-sum rule. In \Cref{fig:bethe_fulfilled} panel (a), we depict the first frequency moment \Cref{eq:f_sum_rule} for a single hydrogen atom. The gray dashed curve depicts the correct analytical dispersion $q^2 /2$. In the solid red line, we show the first moment stemming from the bound-bound transitions, while the green curve depicts the first moment of the bound-free transitions. This makes it apparent that the bound-free transitions, even when accounted for exactly, do not obey the $f$-sum rule alone as previously discussed. Focusing now on the first frequency moment of the bound-bound transitions (solid-red), we observe a maximum at around $1.5$ $\invangs$ and a gradual decrease of $\Omega^{(1)}(q)$ towards increasing $q$. To fulfil the Bethe $f$-sum rule, the addition of the first bound-bound and bound-free frequency moment is shown as the connected blue circles. This curve is in perfect agreement with the analytic relation (dotted gray curve). From the theory section of this work, it becomes clear why. The {\BFSR} is based on the summation of a complete basis set of final states, that requires the addition of bound and free states in the case of the chemically inspired models. In panel (b) we exchanged our exact model with the prediction of the IA (pink dashed). The IA alone does not fulfil the correct quadratic dispersion by itself. We also observe that the sum of the IA and the bound-bound moment falls short of exactly fulfilling the {\BFSR} over the whole wave number range. At around $q=1.5$ $\invangs$ the {\BFSR} seems to be fulfilled but is systematically overestimated for larger $q$. Again, this is due to the approximation of the continuum Coulomb states by plane waves. As we have seen in the previous section, the IA agrees with the {\BFSR} by itself for large q. This is also caused by the systematic decrease of the bound-bound transitions with higher $q$, as no momentum can be transferred there. 

In panel (c) we show the absolute deviation of the individual bound-free moments and combined moments from panel (a) and (b) from the quadratic term. The IA still overestimates the $f$-sum rule for larger $q$ (purple dashed), whereas the exact hydrogen-like bound-free model starts to agree with the {\BFSR} for large $q$. Thus, we observe that neither the IA alone nor the IA with bound-bound transitions included (turquoise circles) completely agree with a quadratic dispersion in the first moment. Only the exact treatment (blue circles) will agree with the Bethe $f$-sum rule over the whole $q$-range. Future efforts therefore have to be taken to improve the existing Chihara models at finite temperature with our consistent bound-state treatment in order to comply with the Bethe $f$-sum rule and thus take all final state transitions into account. The extension of this bound-state treatment to finite temperature would exceed the scope of this work and is a dedicated future endeavour.  

We note here that the perfect agreement with the {\BFSR} is only given if enough bound-bound transitions are accounted for. Thus, the adherence to the $f$-sum rule is another constraint that can be utilized to check if enough bound-bound transitions have been taken into account to produce a theoretically sound DSF in forward-fitting procedures.

\subsection{Simulating cold hydrogen XRTS spectra}

We combine the two newly developed models to compute the ground-state inelastic DSF of atomic hydrogen. Throughout this section, hydrogen is treated as an ensemble of ground-state atoms; molecular bonding and solid-state band-structure effects are not included. In this atomic limit, bound-bound transitions have intrinsic line shapes determined by the natural line-widths of the corresponding atomic transitions. For example, the natural line-width of the $2p \rightarrow 1s$ transition is of the order of $10^{-7}\,\mathrm{eV}$. This line-width is negligible on the energy scales considered here. Therefore, the bound-bound contribution to the DSF can be accurately represented by delta functions.

In an XFEL experiment, however, the measured spectrum is not determined by the intrinsic DSF alone, but also depends on the spectral profile of the incident XFEL beam. The spectrum leaving the sample in the direction of the detector is given by the convolution of the DSF with the source function, which can be approximated by a Voigt profile in first order. Since the natural linewidths of the bound-bound transitions are much smaller than the XFEL bandwidth, the convolution of a bound-bound delta peak with the source function simply produces a Voigt profile centred at the corresponding energy transfer,
$\Delta \omega = \omega_f - \omega_i$.
The width of the observed peak is therefore set by the source function rather than by the intrinsic atomic line-width.

\begin{figure}
\centering
\includegraphics[width=\columnwidth]{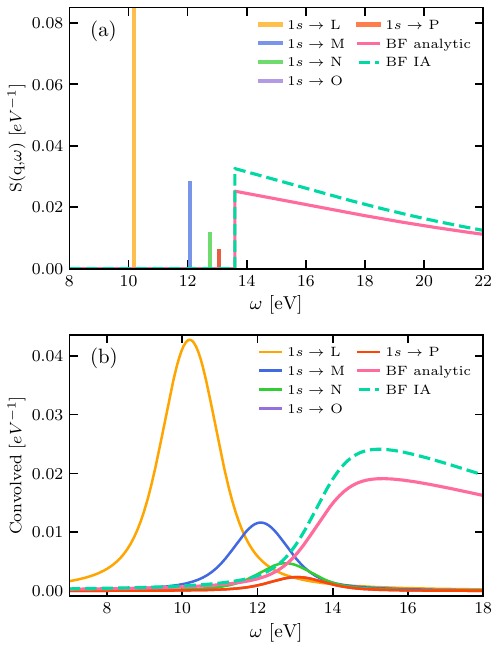}
\caption{Inelastic DSF of ground-state atomic hydrogen. Panel (a) shows the intrinsic DSF, where the bound-bound transitions are represented by delta-like peaks. Panel (b) shows the corresponding source-convolved spectrum obtained by convolution with a Voigt profile using $\sigma = \gamma = 0.5\,\mathrm{eV}$. Spectrum taken at $q = 1.41 \mathrm{\AA}^{-1}$.}
\label{fig:H_jet_spectr_combined}
\end{figure}

This scenario is shown in \Cref{fig:H_jet_spectr_combined}. Panel (a) shows the intrinsic inelastic DSF of ground-state atomic hydrogen for $q=1.41 \mathrm{\AA}^{-1}$. The bound-bound transitions are represented by delta-like lines, while the magenta and turquoise curves show the bound-free contribution obtained from the analytic treatment developed in this work and from the impulse approximation (IA), respectively. The $1s \rightarrow \mathrm{L}$-shell transitions carry by far the largest spectral weight, exceeding both the remaining bound-bound transitions and the bound-free continuum. Panel (b) shows the corresponding source-convolved signal, obtained using a Voigt profile with $\sigma = \gamma = 0.5\,\mathrm{eV}$. After convolution, the dominance of the K- to L-shell bound-bound transitions remains clearly visible. These transitions therefore make a non-negligible contribution to the measured XRTS signal for ground-state atomic hydrogen under the conditions considered here.

To assess the impact of these spectral features on a possible XRTS measurement, we simulate a detector image for the HED end station of the European XFEL \cite{zastrau_high_2021} using the ray-tracing code HEART \cite{gawne_heart_2026}. This allows us to propagate the calculated scattering signal through a realistic experimental geometry and compare the resulting detector response with that obtained from the standard Chihara treatment.

The total XRTS signal also contains an elastic contribution. Within the Chihara decomposition, the Rayleigh weight for a one-component plasma is given by \cite{bellenbaum_x-ray_2026,wunsch_ion_2009}
\begin{equation}
W_R(q) = |f(q) + \phi(q)|^2 S_{ii}(q),    
\end{equation}
where $f(q)$ is the atomic form factor, $\phi(q)$ is the free-electron screening cloud, and $S_{ii}(q)$ is the ion-ion static structure factor. In the present ground-state atomic hydrogen calculation, no free electrons are present. We also assume that the x-ray pulse is sufficiently short that the sample remains unheated during the interaction. Consequently, we set
$\phi(q)=0$.
For the proof-of-concept calculation presented here, we further approximate the ion-ion structure factor by
$S_{ii}(q)=1$.
This approximation neglects correlations in the target and may not be quantitatively accurate at the scattering wave number considered. However, it isolates the effect of the improved bound-bound and bound-free treatment on the simulated XRTS signal. A more realistic calculation can incorporate an appropriate $S_{ii}(q)$ and refined line-shape models without changing the structure of the present approach. 

\begin{figure}
\centering
\includegraphics[width=\columnwidth]{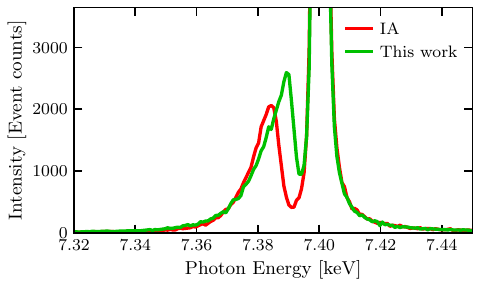}
\caption{Comparison of the simulated XRTS spectra obtained from the standard Chihara model and from the improved hydrogenic model developed in this work under from the situation depicted in \Cref{fig:H_jet_spectr_combined} ($q=1.41 \, \mathrm{\AA}^{-1}$). The improved model includes the bound-bound and bound-free contributions of ground-state atomic hydrogen and therefore predicts additional spectral weight compared with the conventional treatment. The spectrum was obtained using the EuXFEL Jungfrau photon counter \cite{Sikorski_first_2023} preset in the HEART ray-tracing code \cite{gawne_heart_2026}.}
\label{fig:final_comparison}
\end{figure}

The resulting comparison is shown in \Cref{fig:final_comparison} for the same Voigt profile as the previous figure and a beam energy of $7.4 \, \mathrm{keV}$. The standard Chihara model and the improved hydrogenic model lead to visibly different spectral signals. In particular, the improved treatment captures the additional spectral weight associated with bound-bound transitions, which is usually neglected in the conventional description. This demonstrates that bound-state excitations can make a significant contribution to the measured XRTS response of cold, ground-state atomic hydrogen and should be included when interpreting such spectra.

This proof-of-concept calculation shows that bound-bound transitions can play a dominant role even in the simplest possible ground-state atomic system. A true warm-dense-matter system would additionally include elastic and inelastic scattering from free electrons, as well as thermal occupation of excited bound states. A full treatment of the molecular solid-hydrogen phase is beyond the scope of the present work. Future investigations will therefore focus on extending the present chemical model toward a more realistic description of solid hydrogen jet experiments.

\section{\label{sec:conclusion}Conclusion}

In this work, we have provided a formal explanation for why the Bethe $f$-sum rule is systematically violated in the standard Chihara decomposition in the context of XRTS experiments. The primary origin of this violation is the neglect of bound-bound transitions. An additional contribution arises from the use of the impulse approximation (IA), which treats all final states as plane waves and is therefore only asymptotically exact. In particular, elements such as C and Al exhibit non-negligible violations of the Bethe $f$-sum rule even at the high momentum transfers accessible in the backscattering geometry of modern XRTS experiments.

To address this limitation, we used the formal definition of the ground-state DSF and specialized it to hydrogen-like atoms. The key quantity required to extend the Chihara model beyond an IA-only treatment is the exact transition matrix element. We revisited the analytic calculation of hydrogen-like transition matrix elements and adapted these expressions to compute the ground-state DSF. By including both bound-bound transitions and exact bound-free matrix elements, the Bethe $f$-sum rule can be satisfied to arbitrary precision. This provides, to our knowledge, the first implemented bound-state treatment within the Chihara decomposition for XRTS that satisfies this important sum rule. The satisfaction of such sum rules is essential for determining the relative weights of the individual components in finite-temperature Chihara models.

Finally, we demonstrated a practical application of the hydrogenic model to ground-state atomic hydrogen. We showed that the $1s \rightarrow \textrm{L}$-shell transitions contribute non-negligibly to the total DSF. Furthermore, using the HEART ray-tracing code to simulate a detector image at the HED end station of the European XFEL, we demonstrated that the differences introduced by the coherent hydrogen-like treatment should be clearly visible in an experimental measurement.

Future work should focus on extending the model to finite temperatures, thereby enabling application to experimental inference. The present work provides an important basis for such developments because it implements the most computationally expensive calculations analytically and is therefore well suited for adaptation to forward-fitting routines. A second important extension is the inclusion of screening models capable of estimating the electronic structure of multi-electron atoms. To account for the surrounding plasma environment, continuum lowering, or ionization potential depression, should also be incorporated into the hydrogenic model.

In total, we have presented a minimal way to extend the commonly used Chihara decomposition to treat bound states beyond the IA. The model is readily extendable to experimentally relevant situations while retaining a high degree of interpretability. We expect that this approach will enable further developments in XRTS modelling and improve the accuracy of future experimental analyses. The implementation will be made readily available in the \texttt{xDAVE} open source code \cite{bellenbaum_x-ray_2026,xdave_github_repo_v0.1.0}.
\begin{acknowledgments}
The authors acknowledge helpful comments and references provided by Andrew Lanzrath and Brian Wilson.

This work was performed under the auspices of the U.S. Department of Energy by Lawrence Livermore National Laboratory under Contract No. DE-AC52-07NA27344. M.P.B.,~V.A.K.~and P.H.~were supported by the Laboratory Directed Research and Development (LDRD) Grant No.~25-ERD-047. Computing support for this work came from the Lawrence Livermore National Laboratory (LLNL) Institutional Computing Grand Challenge program.

This work has received funding from the European Research Council (ERC) under the European Union’s Horizon 2022 research and innovation programme (Grant agreement No. 101076233, "PREXTREME"). 
Views and opinions expressed are however those of the authors only and do not necessarily reflect those of the European Union or European Research Council Executive Agency. Neither the European Union nor the granting authority can be held responsible for them. This work has received funding from the European Union's Just Transition Fund (JTF) within the project \emph{R\"ontgenlaser-Optimierung der Laserfusion} (ROLF), contract number 5086999001, co-financed by the Saxon state government out of the State budget approved by the Saxon State Parliament. This project has received funding from the Fusion2024 program of the German Federal Ministry of Research, Technology and Space (BMFTR) via the project "VANLIFE" (funding no 13F1016B).

T.D.~gratefully acknowledges funding from the Deutsche Forschungsgemeinschaft (DFG) via project DO 2670/1-1.

\end{acknowledgments}

\appendix
\begin{widetext}

\section{The impulse approximation}\label{app:impulse}

To treat the bound-inelastic scattering contribution in XRTS, practical implementations of the Chihara decomposition often retain only bound-free transitions, which are evaluated using a plane-wave form-factor approximation and, at large momentum transfer, the impulse approximation \cite{eisenberger_compton_1970,schumacher_incoherent_1975}. The bound-free matrix element from \Cref{eq:dsf_bf_nlm} is thus approximated as
\begin{equation}
   M^{bf,IA}_{\mbf{k},n\ell m}(\mbf{q}) = \int\,\mathrm{d}^3r \, {\psi_{\mbf{k}}^\text{free}}^*(\mbf{r}) e^{i\mbf{qr}} \psi^\text{bound}_{n\ell m}(\mbf{r}) \approx 
 \int d^3r \, e^{-i\mbf{k}\cdot\mbf{r}} e^{i\mbf{q}\cdot\mbf{r}} \psi^\text{bound}_{n\ell m}(\mbf{r}) ,
\end{equation}
where the final continuum state, Eq.~\eqref{eq:continuum}, has been replaced by a plane wave $\psi^\text{free}_{\mbf{k}}(\mbf{r}) \approx e^{i\mbf{k}\mbf{r}}$. This reduces the transition matrix element to the Fourier transform of the initial bound orbital evaluated at the momentum difference \(\mbf{p}=\mbf{q}-\mbf{k}\). The impulse approximation now boils down to calculating the bound-free profile by assuming that $k \approx q \gg p$. Thus, it is a good approximation in the limit of large $q$ as indicated by \Cref{fig:violation_higher_Z}. A detailed review of the impulse approximation can be found in Ref.~\cite{chew_impulse_1952}.

\section{\label{app:matrix}Evaluation of the plane-wave matrix element in spherical coordinates} \label{app:bf_numerical}
Using spherical coordinates both bound and continuum wave functions can be separated into a radial and angular part:
$$\psi^\text{bound}_{n\ell m} = R^\text{bound}_{n\ell}(r) Y_{\ell}^m(\theta,\varphi) \qquad \text{and} \qquad \psi^\text{free}_{k\ell m} = R^\text{free}_{k\ell}(r) Y_{\ell}^m(\theta,\varphi),$$
where $Y^m_\ell$ are spherical harmonics.

Following the convention given in Ref.~\cite{pollock_properties_1988}, we use Hartree atomic units and express everything in terms of the parameters:
$$ \kappa = \frac{1}{2\mu} \qquad \text{and} \qquad Z = \frac{Z_1 Z_2}{\kappa},$$
where $\mu$ is the reduced mass between the electron and the nucleus and $Z_1,Z_2$ are the electron and nuclear charge. 
The bound state eigenfunction is given by 
\begin{equation} \label{eq:bound_radial_state}
    R^\text{bound}_{n\ell}(r) = \mathfrak{C}_n^\ell \exp\left(-\frac{r |Z|}{2n} \right) r^\ell L^{2\ell+1}_{n-\ell-1}\left(\frac{r|Z|}{n}\right),
\end{equation}
with normalization constant 
\begin{equation}
    \mathfrak{C}_{n}^\ell = \left(  \frac{|Z|}{n}\right)^{\ell+3/2} \sqrt{\frac{(n-\ell-1)!}{(2n)(n+\ell)!}}.
\end{equation}
We note that the original reference \cite{pollock_properties_1988} contains a small mistake and writes $(n-1)$ in the denominator instead of $(n-\ell)$. 

The continuum Coulomb states are given by 
\begin{equation}
    R^\text{free}_{k\ell}(r) = \mathcal{C}_{k}^\ell e^{-ikr} (2kr)^\ell \ {}_1F_1 \left(\ell+1 - \frac{iZ}{2k}, 2\ell+2; 2ikr\right),
\end{equation}
with the normalization given by 
\begin{equation}
    \mathcal{C}_k^\ell = \sqrt{\frac{2}{\pi}} k e^{-\pi Z /4k} \frac{|\Gamma(\ell+1-iZ/2k)|}{\Gamma(2\ell+1)},
\end{equation}
where $\Gamma(x)$ denotes the Gamma function and ${}_1F_1(\rho,\eta,z)$ denotes the confluent hypergeometric function \cite{bethe_quantum_2008}.

Aligning the axes such that $\vec{r} || \vec{e}_z$, one obtains for the dipole matrix element:
$$
d_{ij} = \bra{j} r \cos\theta\ket{i} = \int\limits_0^\infty dr\, r^3  R_j(r) R_i(r)  \int d\Omega\, {Y_{j}}^*(\Omega) Y_{i}(\Omega) Y_1^0(\Omega) \sqrt{4\pi/3}
$$
after identifying $ \cos\theta =Y_1^0(\Omega) \sqrt{4\pi/3}$.

The angular integral can be directly evaluated using the Wigner-Eckart theorem:
$$\int d\Omega\, {Y_{\ell'}^{m'}}^*(\Omega) Y_{\ell}^m (\Omega) Y_L^0(\Omega) =  (-1)^m \sqrt{\frac{(2L+1)(2\ell' + 1)(2\ell +1)}{4\pi}} \begin{pmatrix} L & \ell & \ell' \\ 0 & 0 & 0  \end{pmatrix} \begin{pmatrix} L & \ell & \ell' \\ 0 & m & -m'  \end{pmatrix}.$$
The Wigner-3j symbols vanish unless $m=m'$ and $\ell+\ell' + 1 = 0$, which are the selection rules for dipole transitions.

In order to obtain a similar expression for the plane-wave matrix element $M_{ij}(q) = \bra{i} e^{i\vec{q}\cdot\vec{r}} \ket{j}$ we insert the expansion
$$e^{iqr\cos\theta} = \sqrt{4\pi} \sum\limits_{L=0}^{\infty} \sqrt{2L+1} i^L J_L(qr) Y_{L}^0 (\theta,\varphi),$$
where $J_L(z)$ is the $L$-th Bessel function of the first kind. One finds:
\begin{equation}\label{eq:mfi_app}
    M_{ij}(q) = \sum\limits_{L=0}^\infty \int\limits_0^\infty dr\, r^2  R_i(r) R_j(r) J_L(qr)   (-1)^m  i^L (2L+1) \sqrt{(2\ell+1)(2\ell'+1)} \begin{pmatrix} L & \ell & \ell' \\ 0 & 0 & 0  \end{pmatrix} \begin{pmatrix} L & \ell & \ell' \\ 0 & m & -m'  \end{pmatrix}.
\end{equation}
This matrix element vanishes unless $m=m'$. However, due to the infinite sum over $L$, there are now contributions for all possible final angular momentum quantum numbers $\ell'$.

While in principle, Eq.~\eqref{eq:mfi_app} gives us everything we need to compute all contributions to the dynamic structure factor, one still needs to carry out the radial integral, which may be done numerically. However, when computing $S^\text{bf}(q,\omega)$ the convergence of the infinite sum over angular momentum quantum numbers $\ell'$ of the continuum states is prohibitively slow, such that practical forward fits would become impossible to carry out in reasonable time. As it turns out, it is also possible to find analytic expressions.

\section{\label{app:bound-bound}Bound-bound matrix element}
Generalizing the result for an $1s$-electron obtained in Ref.~\cite{schnaidt1934}, we can find a closed-form expression for arbitrary bound-bound transitions using parabolic coordinates, which in terms of the Cartesian coordinates are defined as:
$$ \xi = r + z, \quad \eta = r - z, \quad \varphi = \arctan \frac{y}{x}, \quad r = \sqrt{x^2 + y^2 + z^2},$$
where $\xi \in [0,\infty]$, $\eta \in [0,\infty], \varphi \in [0,2\pi]$.
These coordinates provide an alternative way of separating the Schrödinger equation. Instead of having a single orbital quantum number $\ell$, one obtains two quantum numbers $n_1, n_2$, which add up to the usual principal quantum number $n = n_1 + n_2 + m + 1$ \cite{bethe_quantum_2008}. Solutions are of the form:
\begin{equation}\label{eq:psi_para}
\psi^\text{bound}_{n_1n_2m}(\xi,\eta,\varphi) = c  e^{-\frac{1}{2}\frac{Z}{n}(\xi+\eta)} (\xi\eta)^{m/2} (Z/n)^m L_{n_1+m}^m\left(\frac{Z}{n}\xi\right) L_{n_2+m}^m\left(\frac{Z}{n}\eta\right) \frac{e^{im\varphi}}{\sqrt{2\pi}}
\end{equation}
where $L_n^{(\alpha)}(x)$ are the generalized Laguerre polynomials and the normalization constant is given by $$c = \left(\frac{Z}{n}\right)^{m+3/2} \sqrt{\frac{n_1! n_2!}{(n_1+m)!(n_2+m)!}}.$$
A state defined using the quantum numbers $\ket{n\ell m}$ can be expressed as a superposition of parabolic states $\ket{n_1 n_2 m}$:
$$\ket{n\ell m} = \sum\limits_{n_1+n_2+m+1 = n}\braket{n_1 n_2 m}{n\ell m} \ket{n_1 n_2 m},$$
where the overlap integral is given by the Clebsch-Gordan coefficients:
$$\braket{n_1 n_2 m}{n\ell m} = {(-1)^{\frac{n-\ell-1}{2}} \sqrt{2\ell + 1} \begin{pmatrix} \frac{n-1-m}{2} & \frac{n-1+m}{2} & \ell \\ \frac{n_1-n_2}{2} & - \frac{n_1 - n_2}{2} & 0 \end{pmatrix} }.
$$
For $ \vec{q} || \vec{e}_z$ the scattering matrix element for bound states $(n_1,n_2,m)$ and $(n_1',n_2',m')$ is given by:
\begin{equation}
    M(q) = \int\limits_0^{2\pi} d\varphi \frac{e^{i(m-m')\varphi}}{{2\pi}} \int\limits_0^{\infty} d\xi  \int\limits_0^{\infty} d\eta \frac{\xi+\eta}{4} {\psi^\text{bound}_{n_1'n_2'm'}}^\ast \psi^\text{bound}_{n_1n_2m} e^{iq(\xi-\eta)/2}.
\end{equation}
The first integral yields $1$ if $m=m'$ and vanishes otherwise. After inserting Eq.~\eqref{eq:psi_para} one obtains:
\begin{multline*}
M (q) = \frac{cc'}{4}\left(\frac{Z^2}{nn'}\right)^m \int\limits_0^\infty d\xi\int\limits_0^\infty d\eta\, (\xi+\eta) (\xi\eta)^m e^{-\sigma (\xi+\eta) } e^{iq(\xi-\eta)/2} \\ L_{n_1+m}^m\left(\frac{Z}{n}\xi\right) L_{n_2+m}^m\left(\frac{Z}{n}\eta\right) L_{n_1'+m}^m\left(\frac{Z}{n'}\xi\right) L_{n_2'+m}^m\left(\frac{Z}{n'}\eta\right) 
\end{multline*}
where $\sigma = \frac{Z}{2n} + \frac{Z}{2n'}$. Following Ref.~\cite{schnaidt1934}, inserting a partial derivative allows us to get rid of the term $\xi+\eta$, which would otherwise prevent us from separating the coordinates:
\begin{multline*}
M(q) = -\frac{cc'}{4}\left(\frac{Z^2}{nn'}\right)^m \frac{\partial}{\partial\sigma}\int\limits_0^\infty d\xi\int\limits_0^\infty d\eta\, (\xi\eta)^m e^{-\sigma (\xi+\eta) } e^{iq(\xi-\eta)/2} \\ L_{n_1+m}^m\left(\frac{Z}{n}\xi\right) L_{n_2+m}^m\left(\frac{Z}{n}\eta\right) L_{n_1'+m}^m\left(\frac{Z}{n'}\xi\right) L_{n_2'+m}^m\left(\frac{Z}{n'}\eta\right) .
\end{multline*}
The remaining task at hand is to solve the integral:
$$
I_\pm = \int\limits_0^\infty d\eta\, \eta^m e^{-(\sigma\pm iq/2)\eta} L_{n_2+m}^m\left(\frac{Z}{n}\eta\right)  L_{n_2'+m}^m\left(\frac{Z}{n'}\eta\right).
$$
Inserting the explicit expression for the Laguerre polynomials:
$$
L_n^\alpha = \sum\limits_{k=0}^{n} (-1)^k \binom{n+\alpha}{n-k} \frac{x^k}{k!} 
$$
leaves us with:
$$
I_\pm = \sum\limits_{k=0}^{n_2+m} \sum\limits_{j=0}^{n_2'+m} (-1)^{k+j} \binom{n_2 + 2m}{n_2 + m - k} \binom{n_2' + 2m}{n_2 + m - j} \int\limits_0^\infty d\eta\, \eta^m \frac{( Z\eta / n)^k}{k!} \frac{(Z\eta / n')^j}{j!} e^{-(\sigma \pm iq/2)\eta},
$$
which using
$\int_0^\infty d\eta\, \eta^m e^{-a\eta} = m!/a^{m+1}$
can be evaluated, yielding:
$$ I_\pm = \sum\limits_{k=0}^{n_2+m} \sum\limits_{j=0}^{n_2'+m} \frac{(-1)^{k+j}}{k!j!} \binom{n_2 + 2m}{n_2 + m - k} \binom{n_2' + 2m}{n_2' + m - j} \left(\frac{Z}{n}\right)^k \left(\frac{Z}{n'}\right)^j \frac{(k+j+m)!}{(\sigma \pm iq/2)^{k+j+m+1}}.$$
The final expression for the scattering matrix element is given by:
$$
M(q) = -\frac{cc'}{4} \left(\frac{Z}{nn'}\right)^m \frac{\partial}{\partial\sigma} \left[ I_+^{n_2,n_2'} I_-^{n_1,n_1'} \right]. 
$$

\section{Bound-free matrix element} \label{app:bound-free}
To solve the bound-free matrix element one first writes down the free Coulomb wave function \cite{bethe_quantum_2008} in terms of the parabolic coordinates
\begin{align}
    \xi = r + \mathbf{r} \cdot \hat{k}, &\phantom{=}& \zeta = r - \mathbf{r} \cdot \hat{k}, &\phantom{=}& \hat{k} = \vec{k}/k,
\end{align}
such that 
\begin{equation}
    \Psi_{\mbf{k}}(\mbf{r}) = \Gamma(1-i\eta)e^{-\pi\eta/2} e^{i\mbf{kr}} {}_1F_1(i\eta,1,-ikr - i \mbf{k} \cdot \mbf{r}),
\end{equation}
with $\eta = Z/ k$, where $Z$ is given by the core charge of the nucleus \cite{Landau:1991wop,nordsieck_reduction_1954}.
We now write the bound-free matrix element integral as 
\begin{equation}\label{eq:bound-free}
    M^\text{bf}_{\mathbf{k},n\ell m}(\mbf{q}) = e^{-\pi\eta/2} \Gamma(1+i\eta) \int \mathrm{d}^3\mbf{r} \, e^{i\mbf{pr}}{}_1F_1(-i\eta;1;i(kr + \mbf{k}\cdot\mbf{r})) \psi_{n\ell m}(\mbf{r}),
\end{equation}
with define $\mbf{p} = \mbf{q} - \mbf{k}$. We now follow the idea of \cite{nordsieck_reduction_1954,bethe_theory_1954,belkic_bound-free_1981} in a simplified form using two observations. First we look at the analytical solution of the Nordsieck \cite{nordsieck_reduction_1954,bethe_theory_1954} integral for $a_1 =0$ and $\mbf{p}_1 = \mbf{0}$ 
\begin{equation}\label{eq:base_analytical}
    I_0 = \int \mathrm{d}\mbf{r} \,  \, \frac{\e^{i\mbf{pr}-\lambda r}}{r} {}_1F_1(ia,1,ikr - i \mbf{kr}) = \frac{4\pi}{\mbf{p}^2 + \lambda^2} \left( \frac{\mbf{p}^2 + \lambda^2}{\mbf{p}^2 + \lambda^2  + 2\mathbf{kp} - i 2\lambda k} \right)^{ia}.
\end{equation}
This result can purely be expressed in terms of (inserting $\mbf{p} = \mbf{q - k}$)
\begin{equation}
    A =|\mbf{p}|^2 + \lambda^2 = q^2 + k^2 - 2 kq \underbrace{\cos(\vartheta)}_{\mu} + \lambda^2
\end{equation}
and 
\begin{align}
    B &= A + 2 \mbf{kp} - 2 i \lambda k \\
    &= q^2 - k^2 + \lambda^2 + 2i\lambda k.
\end{align}
The total solution then simply becomes a function of the angle between $\mbf{q}$ and $\mbf{k}$ such that
\begin{align}
I_{0}(k,q,\lambda,\mu;a) &= \frac{4\pi}{A} \left( \frac{A}{B }\right)^{ia}     \\
&= \frac{4 \pi}{q^2-k^2 - 2 k q \mu + \lambda^2} \left( \frac{q^2 + k^2 - 2 kq \mu + \lambda^2}{q^2 - k^2 + \lambda^2 - 2 i \lambda k} \right). \label{eq:base_expr}
\end{align}
Since the expression has no dependency on the azimuthal component, one can set $k_y = 0$ for implementation purposes.

The first idea is to reduce this expression, as done in Ref.~\cite{bethe_theory_1954}, to a fundamental integral by setting $a = \eta$ and $\lambda  = \varepsilon_n$, where $\varepsilon_n$ is the bound-state energy, by calculating the derivative 
\begin{equation}\label{eq:base_integral}
    I_1 = -\frac{\partial}{\partial \lambda} I_0 = \int d^3r \ r \ \e^{i\mathbf{pr} - \lambda r} {}_1F_1(ia,1,ikr - i \mbf{kr}).
\end{equation}
The second main idea is to see that the exponential function in \Cref{eq:base_integral} can be utilized as a generating function for both the Laguerre and Laguerre polynomials in \Cref{eq:bound_radial_state} and the Legendre polynomials in the spherical harmonics $Y_{\ell}^m(\vartheta,\varphi)$. We define
$\delta = \frac{|Z|}{2n}$ and rewrite the total wave-function as given in the system of units from \Cref{app:bf_numerical}
\begin{equation}
    \psi_{n\ell m}(\mbf{r}) = \left( 2 \delta\right)^{\frac{3}{2}} \sqrt{\frac{(n-\ell - 1)!}{(2n)(n+\ell)!}} e^{\delta r} L^{2\ell +1}_{n-\ell -1}(2\delta r)^\ell Y_{\ell}^m(\vartheta,\varphi).
\end{equation}
The first step is to use the explicit form of the Laguerre polynomials 
\begin{equation}
    L^{2\ell+ 1}_{n-\ell-1}(2 \delta r) = \sum_{t=0}^{n-\ell-1} (-1)^t \binom{n+\ell}{n - \ell- 1 - t} \frac{1}{t!} (2 \delta)^t r^t, 
\end{equation}
where the $r^t$ term can be generated, similar to the idea of the previous chapter, by 
\begin{equation} \label{eq:I1_generator}
    \int d^3\mathbf{r} \, r^t \, \e^{i\mbf{pr} - \lambda r} {}_1F_1(ia, 1, ikr - i\mbf{kr}) = \left(- \frac{\partial}{\partial \lambda}\right)^t I_1,
\end{equation}
thus the Laguerre polynomials can easily be generated as a sum over derivatives of the base integral expression. 
The next step is to use the solid-harmonics defined as \cite{ribaldone_spherical_2025} 
\begin{equation}
\mathcal{Y}_{\ell m}(\mathbf{r}) = r^\ell Y_{\ell}^{m}(\vartheta,\varphi),
\end{equation}
and using the Cartesian representation of the solid harmonics as indicated in Ref.~\cite[p.133]{varshalovich1988quantum}
\begin{equation}
    \mathcal{Y}_{\ell m}(\mathbf{r}) = \underbrace{\sqrt{\frac{2\ell+1}{4\pi}(\ell +m)!(\ell-m)!}}_{=N_{\ell m}}\sum_{p,q,b} \frac{1}{p!q!b!} \left(-\frac{x + iy}{2} \right)^p \left( \frac{x - iy}{2}\right)^q z^b,
\end{equation}
where $p,q,b$ are the integers fulfilling the conditions $p+q+r = l$ and $p-q = m$. 
For explicit computations it is more useful to choose the following expression with the additional phase convention for the solid harmonic
\begin{equation}
    \mathcal{Y}_{\ell m}(\mbf{r}) = N_{\ell m} r^\ell \e^{im\varphi} (-1)^m \sin^{|m|}(\vartheta) \frac{d^{|m|}}{d\cos(\vartheta)^{|m|}} \frac{1}{2^\ell \ell!}\frac{d^\ell}{d\cos(\vartheta)^\ell} (\cos(\vartheta)^2 - 1)^\ell.
\end{equation}
Using the explicit Rodriguez formula one can obtain
\begin{align} \label{eq:Rodruigez}
    \frac{1}{2^\ell \ell!}\frac{d^\ell}{d\cos(\vartheta)^\ell} (\cos(\vartheta)^2 - 1)^\ell = \frac{1
    }{2^\ell}\sum_{g=0}^{\lfloor \ell /2\rfloor} (-1)^g \binom{\ell}{n} \binom{2\ell - 2g}{\ell} (\cos(\vartheta))^{\ell-2g}.
\end{align}
Taking the additional m-th derivative of this expression results in only the survival of the terms where $ \ell - 2g + 1 > |m|$. We briefly rewrite the term using $u = \cos(\vartheta)$
\begin{equation}
    \frac{d^{|m|}}{du^{|m|}} P_\ell(u) = \sum_{\substack{b=0 \\ (\ell - |m| - b \text{ even})}}^{\ell - |m|} g_b u^b.
\end{equation}
Instead of an explicit formula for $g_b$, it is more practical to first implement the polynomial in \Cref{eq:Rodruigez} and compute the corresponding indices using e.g.~numpy \textit{polyder} function \cite{harris2020array}. Using $\cos(\vartheta) = z/r$ and \cite{ribaldone_spherical_2025}
\begin{equation}
    r^{m} \sin^{|m|}(\vartheta)e^{is|m|} = (x + i s y),
\end{equation}
with $s = m / |m|$ as the sign of m. We then write down the solid-harmonic in Cartesian coordinates 
\begin{equation}
    \mathcal{Y}_{\ell m}(\mathbf{r}) = N_{\ell m} (x + isy)^{|m|} \frac{1}{2^\ell} \sum_{b=0}^{\ell - |m|} g_b z^b r^{\ell - |m|- b}.
\end{equation}
The next step is coming back to \Cref{eq:I1_generator}, is to use the exponential in the integral to generate arbitrary power terms of each coordinate using
\begin{equation}
    (-i)^{|m|} \left(\frac{\partial}{\partial q_x} + i s \frac{\partial}{\partial q_y}\right)^{|m|} I_1 = \int \mathrm{d}^3r (x + isy)^{|m|} e^{i\mbf{pr-\lambda r}} {}_1F_1(ia, 1, ikr - {\mbf{ikr}}).
\end{equation}
This can be expanded in an analogous way to generate arbitrary powers for $z$ by using 
\begin{equation}
     (-i)^{|m|} \left(\frac{\partial}{\partial q_z}\right)^{|m|} I_1 = \int \mathrm{d}^3r \ z^{|m|} \ e^{i\mbf{pr-\lambda r}} {}_1F_1(ia, 1, ikr - {\mbf{ikr}}).
\end{equation}
Thus, all the derivatives enable one to formulate the bound-free matrix element as a sum over derivatives of the base integral \Cref{eq:base_integral} with the known coefficients for the polynomial terms involved. 
In total the bound-free matrix element is then given by for the case of $\hat{q} || \hat{z}$:
\begin{align}
M^\text{bf}_{\mathbf{k},n\ell m}(\mathbf{q}) &= \mathfrak{C}_{n\ell} N_{\ell m} (-1)^m \e^{-\pi\eta/2} \Gamma(1+i\eta) \sum_{t=0}^{n-\ell-1}\sum_{\substack{b=0 \\ (\ell-|m|-b) \text{ even}}}^{\ell-|m|} (-1)^t\binom{n+\ell}{n-\ell-1-t}\frac{(2\delta)^t}{t!}\cdot g_b  \label{eq:bound_free_analytic} \\ \nonumber
&\times \left(-\frac{\partial}{\partial\lambda}\right)^{t+\ell-|m|-b+1} \left(-i\frac{\partial}{\partial q_z}\right)^b (-i)^{|m|}\left(\frac{\partial}{\partial q_x}+is\frac{\partial}{\partial q_y}\right)^{|m|} I_0\Bigg|_{\substack{\lambda=\delta \ q_x=q_y=0}}. 
\end{align}
This expression can be computed by implementing the analytical expression \cref{eq:base_analytical} in a computer algebra system such as SymPy \cite{10.7717/peerj-cs.103} and evaluating all computational derivatives there. The resulting expression is then simply a sum over analytical expression that can be evaluated very efficiently and is therefore suitable for practical fitting purposes.
To verify out our approach with respect to the dynamic structure factor, we can simply calculate the bound-free dynamic structure factor by using the fact the base expression in \Cref{eq:base_expr} only depends on the azimuthal component $\mu \in [-1,1]$. Thus we simply have to integrate over the $\mu$ dependence in the expression of the final result. We show here the comparison of the analytic expression of the bound-free DSF versus the analytic result for the state $\ket{3,2,-1}$ in \Cref{fig:bf_numerical_analytic}.
\begin{figure}[h]
    \centering
    \includegraphics[width=0.5\linewidth]{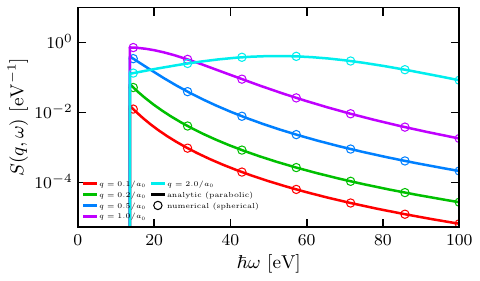}
    \caption{Comparison of the bound-free dynamic structure factor of hydrogen in the ground state for the state $\ket{1,0,0}$. The numerical solution took a total runtime of $2297.6$ $s$ with 8 points per curve, while the analytical solution could be calculated in $0.7$ $s$ for 500 points per curve.}
    \label{fig:bf_numerical_analytic}
\end{figure}
The figure shows the numerical solution as the circles for various values of $q$, while the solid lines depict the analytic solution. The agreement is excellent and the runtime of the analytic solution offered a speed-up of the a factor of 3282.28 (8 points per curve numerical, 500 points per curve analytic). This makes this technique practical for fitting real-world spectra.
\section{Dipole limit, TRK sum rule}

In the long-wavelength limit ($q \to 0$), the transition matrix element $M_{ij}(\qv)$ reduces to that of the dipole operator
\begin{equation}\label{eq:dipole_limit}
    |M_{ij}(\qv)|^2 = |\bra{i} e^{i\mbf{qr}} \ket{j}|^2 \approx \delta_{ij} + q^2 |\underbrace{\bra{i} \hat{\mbf{r}} \ket{j}}_{d_{ij}}|^2.
\end{equation}
While for any transition $i\to j$, there is always some $q>0$ at which $M_{ij}$ assumes a finite value, the dipole matrix element $d_{ij}$ vanishes except for transitions which are \textit{dipole-allowed}, which are exactly those in which the angular momentum quantum number changes by one, $\Delta \ell = \pm 1$, and the magnetic quantum number is conserved, $\Delta m = 0$ (see App.~\ref{app:bound-bound} for a discussion of the selection rules).

The Thomas-Reiche-Kuhn (TRK) or $f$-sum rule states that for any given initial state $i$ the oscillator strengths $f_{ij}$ of all possible transitions $i \to j$ must add up to one
\begin{equation}\label{eq:trk}
\frac{2m}{\hbar^2} \sum\limits_{j} (\varepsilon_j-\varepsilon_i)|\underbrace{\bra{i} \hat{\mbf{r}} \ket{j}}_{d_{ij}} |^2 = \sum\limits_j f_{ij} = 1.
\end{equation}
Eq.~\eqref{eq:trk} therefore gives the normalization condition for the photoabsorption/dipole spectrum:
\begin{equation}\label{eq:conductivity}
\operatorname{Re} \sigma(\omega) = \sum\limits_{ij} |\bra{j} \hat{\mbf{r}} \ket{i} |^2 \, \omega\,  \delta(\omega - \varepsilon_j + \varepsilon_i),
\end{equation}
which describes the optical (no momentum transfer) limit of the DSF.
The magnitude of different contributions to the TRK $f$-sum rule when splitting the sum into bound states and continuum is well known in atomic physics.
For example,  Eq.~\eqref{eq:conductivity} becomes
\begin{multline*}
\operatorname{Re}\sigma(\omega) = \sum_{n\ell m} | \bra{1s} \mbf{r} \ket{n\ell m}|^2 \, \omega\, \delta ( \omega - \varepsilon_{n} + \varepsilon_{1s} ) 
+ \int\limits_0^\infty dk \, | \bra{1s} \mbf{r} \ket{k \ell m}|^2 \, \omega\, \delta ( \omega - \epsilon_k + \varepsilon_{1s} ),
\end{multline*}
for a single electron with $i = \ket{1s}$  \cite{Ogilvie2014}.
Bethe and Salpeter \cite{bethe_quantum_2008} give a partitioning of $0.5650 + 0.4350 = 1$ into bound-bound and bound-free transitions. Evidently, in the long-wavelength limit, one is neglecting the largest contribution to the $f$-sum rule when only taking bound-free transitions into account.

From these preliminary observations, we will give the expression for the DSF of a hydrogen-like atom for both bound-bound and bound-free transitions after a brief introduction to the Chihara decomposition. These analytical relations have been well known and documented \cite{belkic_bound-free_1981,moses_bounds_1994}, however they have, to the best of out knowledge, never been used or implemented in the context of XRTS to analyse experiments. In general, the following models are an extension of the inelastic bound-state treatment beyond only the impulse approximation (IA).

\end{widetext}

\bibliographystyle{apsrev4-2}
\bibliography{references}  


\end{document}